\documentclass[11 pt, a4 paper,]{article}
\pdfoutput=1
\usepackage[english]{babel}
\usepackage[utf8]{inputenc}
\usepackage{graphicx}
\usepackage{amssymb}
\usepackage{dcolumn}
\usepackage{bm}
\usepackage{color}
\usepackage[colorlinks=true, linkcolor=blue, citecolor=blue]{hyperref}
\usepackage{subfigure}
\usepackage{graphicx,psfrag,xspace}
\usepackage{amssymb,amsfonts,amsmath}
\usepackage{setspace}
\usepackage{latexsym}
\usepackage{vmargin}
\usepackage{lmodern}
\usepackage{amsmath}
\usepackage{amsfonts}
\usepackage{amssymb}
\usepackage{caption}
\usepackage{textcomp}
\usepackage{graphicx}
\usepackage{pstricks}
\usepackage{mathtools}
\usepackage{hyperref}
\usepackage{wrapfig}
\usepackage{float}
\usepackage{authblk}
\usepackage[left=2cm,right=2cm,top=1.5cm,bottom=2cm,includeheadfoot,a4paper]{geometry}

\DeclarePairedDelimiter{\floor}{\lfloor}{\rfloor}

\newcommand{\lp}{\left(}
\newcommand{\rp}{\right)}
\renewcommand{\d}{\ensuremath{\mathrm{d}}}

\newcommand{\expect}[1]{\mathbb{E}\left[  #1 \right] }
\newcommand{\empiri}[1]{\left\langle  #1 \right\rangle }

\setmarginsrb{30 mm}{20 mm}{30 mm}{20 mm}{0 mm}{10 mm}{0 mm}{10 mm}
\begin{document}
\title{When does inequality freeze an economy?}
\author[1,2]{Jo\~{a}o  Pedro Jerico}
\author[2]{Fran\c{c}ois P. Landes}
\author[2]{Matteo Marsili}
\author[3]{Isaac P\'erez Castillo}
\author[4,5]{Valerio Volpati}

\affil[1]{Departamento de Fisica Geral, Instituto de Fisica, Universidade de S\~ao Paulo, CP 66318, 05315-970, S\~ao Paulo-SP, Brazil} 
\affil[2]{The Abdus Salam International Centre For Theoretical Physics (ICTP), Strada Costiera 11, 34151 Trieste, Italy}
\affil[3]{Department of Complex Systems, Institute of Physics, UNAM, P.O. Box 20-364, 01000 Cd. Mx., M\'exico}
\affil[4]{International School for Advanced Studies (SISSA), Via Bonomea 265, 34136 Trieste, Italy}
\affil[5]{INFN, Sezione di Trieste, Via Valerio 2, 34126, Trieste Italy}

\date{\today}

\maketitle
  
\begin{abstract}
Inequality and its consequences are the  subject of intense recent debate.  Using a simplified model of the economy, we address the relation between inequality and liquidity, the latter understood as the frequency of economic exchanges. Assuming a Pareto distribution of wealth for the agents, that is consistent with empirical findings, we find an inverse relation between wealth inequality and overall liquidity. 
We show that an increase in the inequality of wealth results in an even sharper concentration of the liquid financial resources. 
This  leads to  a congestion of the flow of goods and the arrest of the economy when the Pareto exponent reaches one.
\end{abstract}

\section{Introduction}
Today's global economy is more interconnected and complex than ever, and seems out of any particular institution's control. The diversity of markets and traded products, the complexity of their structure and regulation, make it a daunting challenge to understand behaviours, predict trends or prevent systemic crises. The neo-classical approach, that aimed at explaining global behaviour in terms of perfectly rational actors, has largely failed \cite{SMD30,BouchaudCrisis,KirmanBook}. Yet, persistent statistical regularities in empirical data suggest that a less ambitious goal of explaining economic phenomena as emergent statistical properties of a large interacting system may be possible, without requiring much from agents' rationality (see e.g. \cite{Gode1993,Smith2003}). One of the most robust empirical stylised fact, since the work of Pareto, is the observation of a broad distribution of wealth which approximately follows a power law.  Such a power law distribution of wealth does not require sophisticated assumptions on the rationality of players, but it can be reproduced by a plethora of simple models (see e.g. \cite{bouchaud2000wealth,Yakovenko2009Review,Gabaix2009,Sornette2013}), in which it emerges as a typical behaviour -- i.e. as the behaviour that the system exhibits with very high probability -- within quite generic settings. 

The debate on inequality has a long history, dating back at least to the work of Kutznets \cite{Kuznets1955} on the u-shaped relationship  of inequality on development. Much research has focused on the relation between inequality and growth (see e.g. \cite{PerssonTabellini1994}). Inequality has also been suggested to be positively correlated with a number of indicators of social disfunction, from infant mortality and health to social mobility and crime \cite{TheSpirit}.

\begin{figure}
\centering
\includegraphics[width=1.0\textwidth]{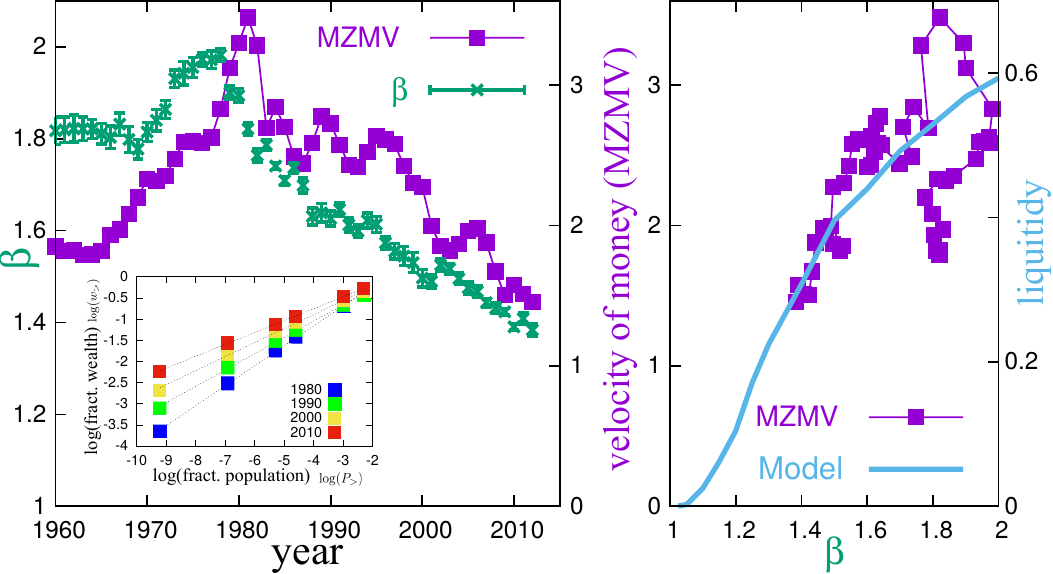}
\caption{Left: Velocity of money of MZM stocks (right y-axis) and Pareto exponent $\beta$ of the wealth distribution (left y-axis) as a function of time. Both time series refer to the US. The data on the money velocity is retrieved from \cite{FRED}, the data on the wealth distribution is taken from \cite{SaezZucman2016}. Inset: relation between the fraction $w_>$ of wealth owned by the $P_>$ percent wealthiest individuals, and $P_>$ for the years 1980, 1990, 2000 and 2010 (see footnote \ref{foot:beta fit}).
Right: MZM velocity of money (MZMV, central y-axis) as a function of $\beta$, for the same data. Liquidity, defined as the probability that a unit-money random exchange takes place,  (right y-axis) as a function of $\beta$, in the synthetic economy described by our model (see Eq. \ref{def:pavg} and Figure \ref{Fig:K10_ps_beta} for details on the numerical simulations).} 
\label{Fig:data}
\end{figure}

The subject has regained much interest recently, in view of the claim that levels of inequality have reached the same levels as in the beginning of the 20th century \cite{Piketty2001}. Saez and Zucman \cite{SaezZucman2016} corroborate these findings, studying the evolution of the distribution of wealth in the US economy over the last century, and they find an increasing concentration of wealth in the hands of the 0.01\% of the richest. Figure \ref{Fig:data} shows that the data in Saez and Zucman \cite{SaezZucman2016} is consistent with a power law distribution $P\{w_i>x\}\sim x^{-\beta}$, with a good agreement down to the 10\% of the richest (see caption\footnote{\label{foot:beta fit}
Ref. \cite{SaezZucman2016} reports the fraction $w_>$ of wealth in the hands of the $P_>=10\%, 5\%, 1\%, 0.5\%, 0.1\%$ and $0.01\%$ richest individuals.  If the fraction of individuals with wealth larger than $w$ is proportional to $P_>(w)\sim w^{-\beta}$, the wealth share $w_>$ in the hands of the richest $P_>$ percent of the population satisfies $w_>\sim P_>^{1-1/\beta}$ (for $\beta>1$). Hence $\beta$ is estimated from the slope of the relation between $\log P_>$ and $\log w_>$, shown in the inset of Fig. \ref{Fig:data} (left) for a few representative years. The error on $\beta$ is computed as three standard deviations in the least square fit. 
}).
 The exponent $\beta$ has been steadily decreasing in the last 30 years,  
reaching the same levels it attained at the beginning of the 20th century ($\beta=1.43\pm 0.01$ in 1917).

Rather than focusing on the determinants of inequality, here we focus on a specific consequence of inequality, i.e. on its impact on liquidity.  There are a number of reasons why this is relevant. First of all, the efficiency of a market economy essentially resides on its ability to allow agents to exchange goods.
A direct measure of the efficiency is the number of possible exchanges that can be realised or equivalently the probability that a random exchange can take place. This probability  quantifies the ``fluidity'' of exchanges and we shall call it {\em liquidity} in what follows. This is the primary measure of efficiency that we shall focus on. 
Secondly, liquidity, as intended here, has  been the primary concern of monetary polices such as Quantitative Easing aimed at contrasting deflation and the slowing down of the economy, in the aftermath of the 2008 financial crisis. A quantitative measure of liquidity is provided by the {\em velocity of money} \cite{IFisher}, measured as the ratio between the nominal Gross Domestic Product and the money stock\footnote{We report data on the MZM (money with zero maturity), the broadest definition of money stock that includes all money market funds. We refer to \cite{FRED} for further details.} and it quantifies how often a unit of currency changes hand within the economy. As Figure \ref{Fig:data} shows, the velocity of money has been steadily declining in the last decades. This paper suggests that this decline and the increasing level of inequality are not a coincidence. Rather the former is a consequence of the latter. 

Without clear yardsticks marking levels of inequality that seriously hamper the functioning of an economy, the debate on inequality runs the risk of remaining at a qualitative or ideological level. Our main finding is that, in the simplified setting of our model, there is a sharp threshold beyond which inequality becomes intolerable. More precisely, when the power law exponent of the wealth distribution approaches one from above, liquidity vanishes and the economy halts because all available (liquid) financial resources concentrate in the hands of few agents. This provides a precise, quantitative measure of when inequality becomes too much. 

Our main goal in the present work is  thus to isolate the relation between inequality and liquidity in the simplest possible model that allows us to draw sharp and robust conclusions.  Specifically, the model is based on a simplified trading dynamics in which agents with a Pareto distributed wealth randomly trade goods of different prices.  Agents receive offers to buy goods and each such transaction is executed if it is compatible with the budget constraint of the buying agent. This reflects a situation where, at those prices, agents are indifferent between all feasible allocations. The model is in the spirit of random exchange models (see e.g. \cite{Foley1994,SjurFlam2012}), but our emphasis is not on whether the equilibrium can be reached or not. In fact we show that the dynamics converges to a steady state, which corresponds to a maximally entropic state where all feasible allocations occur with the same probability. Rather we focus on the allocation of cash in the resulting stationary state and on the liquidity of the economy, defined as the fraction of attempted exchanges that are successful. We remark that since the wealth distribution is fixed, the causal link between inequality and liquidity is clear in the simplified setting we consider.

Within our model, the freezing of the economy occurs because when inequality in the wealth distribution increases, financial resources (i.e. cash) concentrate more and more in the hands of few agents (the wealthiest),
leaving the vast majority without the financial means to trade. This ultimately suppresses the probability of successful exchanges, i.e. liquidity (see Figure \ref{Fig:data}, right). 

This paper is organised as follows: we start by describing the model and its basic characteristics in Section \ref{sec:modelDescription},  providing a quick overview of the main results and features of the model in Section \ref{sec:main_features}.  In  Section \ref{sec:solutionsNumerical_and_analytical} we explain in more detail how these features can be understood by an approximated solution of the Master Equation governing the trading dynamics. Details on the analytical derivations and Monte Carlo simulations are thoroughly presented in the appendices. We conclude with some remarks in Section \ref{sec:con}.

\section{The model}
\label{sec:modelDescription}
The model consists of $N$ agents, each with wealth $c_i$ with $ i = 1, \ldots, N$. Agents are allowed to trade among themselves $M$ objects. Each object $ m = 1, \ldots,M$ has a price $\pi_m$. 
A given allocation of goods among the agents is described by an $N\times M$ allocation matrix $\mathcal{A}$ with entries $a_{i,m} = 1$ if agent $i$ owns good $m$ and zero otherwise. 
Agents can only own baskets of goods that they can afford, i.e. whose total value does not exceed their wealth. The wealth not invested in goods
\begin{equation}
c_i -  \sum_{m=1}^M a_{i , m} \pi_m =\ell_i\ge 0, \quad\quad i=1,\ldots,N,
\label{eq:1}
\end{equation}
corresponds to the cash (liquid capital) that agent $i$ has available for trading. 
The inequality $\ell_i\ge 0$ for all $i$ indicate that lending is not allowed. Therefore 
the set of feasible allocations -- those for which $\ell_i\ge 0$ for all $i$ -- is only a small fraction of the $M^N$ conceivable allocation matrices $\mathcal{A}$.

Starting from a feasible allocation matrix $\mathcal{A}$, we introduce a random trading dynamics in which a good  $m$ is picked uniformly at random among all goods. Its  owner  then attempts to sell it to another agent $i$ drawn uniformly at random among the other agents. If agent $i$ has enough cash to buy the product $m$, that is if $\ell_i \ge \pi_m$, the transaction is successful and his/her cash decreases by $\pi_m$ while the cash of the seller  increases by $\pi_m$. We do not allow  objects to be divided. Notice that the total capital $c_i$ of agents does not change over time, so $c_i$ and the  prices $\pi_m$ are parameters of the model. The entries of the allocation matrix, and consequently the cash, are dynamical variables, which  evolve over time according to this dynamics. 
This model belongs to the class of zero-intelligent agent-based models, in the sense that agents do not try to maximize any utility function.

An interesting property of our dynamics is that the stochastic transition matrix $W(\mathcal{A}\to\mathcal{A}') $ is symmetric between any two feasible configurations $\mathcal{A}$ and $\mathcal{A}'$: $W(\mathcal{A}\to \mathcal{A}')=W(\mathcal{A}'\to \mathcal{A})$. 
We note that any feasible allocation  $\mathcal{A}$ can be reached from any other feasible allocation  $\mathcal{A}'$ by a sequence of trades.
This implies that the dynamics satisfies the detailed balance condition, with a stationary distribution over the space of feasible configurations that is uniform: $P(\mathcal{A})= \text{const}$.  Alternative choices of dynamics which also fulfil these conditions are explored in appendix \ref{app:rules_enumeration}.

In particular, we focus on realisations where the wealth $c_i$ is drawn from a Pareto distribution $P\{c_i> c\} \sim c^{-\beta}$, for $c>c_{\min}$ for each agent $i$. We let $\beta$ vary to explore different levels of inequality, and compare different economies in which the ratio between the total wealth $C=\sum_i c_i$ and the total value of all objects $\Pi=\sum_{m}\pi_m$ is kept fixed. 
We use $C>\Pi$ so as to have feasible allocations.
We consider cases where the $M$ objects are divided into a small number $K$ of 
classes with $M_k$ objects per class ($k=1,\ldots,K$); objects belonging to class $k$ have the same price $\pi_{(k)}$.
If $z_{i,k}$ is the number of object of class $k$ that agent $i$ owns, then \eqref{eq:1} takes the form 
$c_i =  \sum_{k=1}^K z_{i , k} \pi_{(k)} + \ell_i$. 

\section{Main results}
\label{sec:main_features}

The main result of this model is that the flow of goods among agents becomes more and more congested as  inequality increases until it halts completely when the Pareto exponent $\beta$ tends to one from above.

\begin{figure}
\centering
\includegraphics[width=\textwidth]{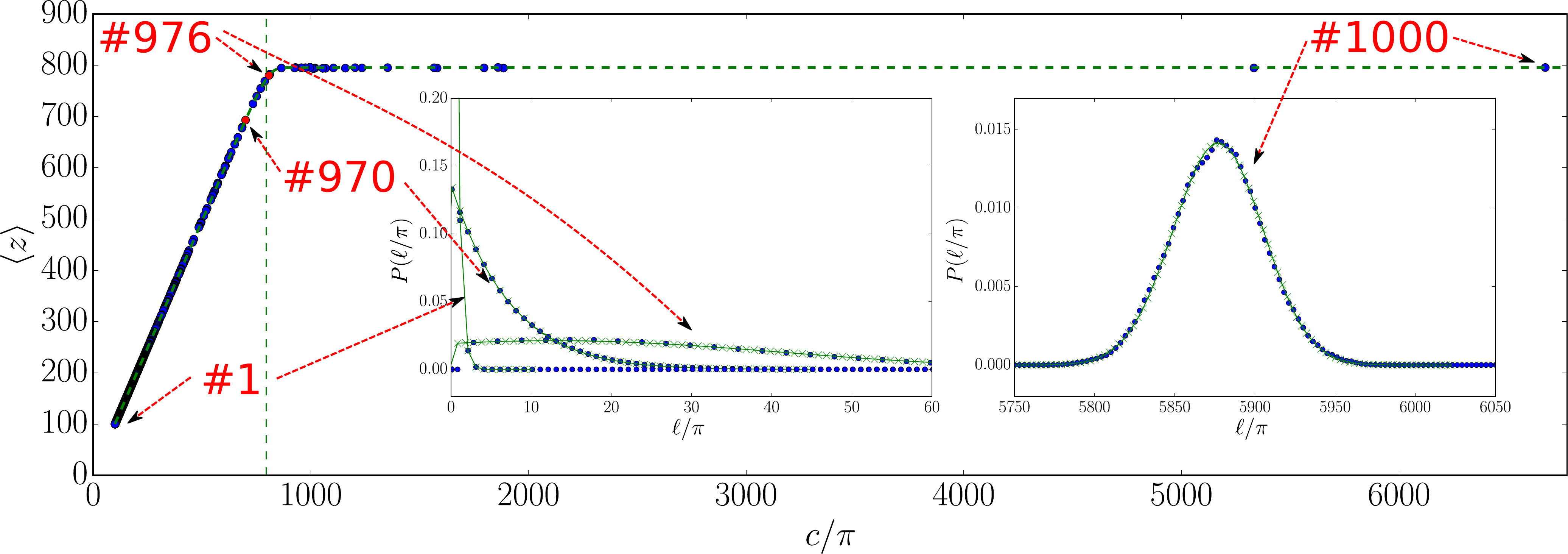}
\caption{Capital composition in an economy with a single type of good, $N=10^3$ agents, $\beta=1.8$, $M\approx 2 . 10^5$ and $C/\Pi=1.1$. Points $\{(\empiri{z}_i,c_i)\}_{i=1}^{N}$ denote the average composition of capital for different agents obtained in Monte Carlo simulations. 
This is compared with the analytical solution obtained from the Master Equation (green dashed line) given by  Eq. (\ref{Eq:ME_solution_1Good}).
The vertical dashed line at $c^{(1)}\simeq 7.98=M/N p^{(\text{suc})}_1$ indicates the analytically predicted value of the crossover wealth that separates the two classes of agents.
Insets: cash distributions $P_i(\ell)$ of the indicated agents. 
}
\label{Fig:Picturesque_RichPoor_transition_beta}
\end{figure}

The origin of this behaviour can be understood in the simplest setting where $K=1$, i.e. all goods have the same price $\pi_m=\pi_{(1)} = \pi$ (we are going to omit the subscript $(1)$ in this case). 
Figure \ref{Fig:Picturesque_RichPoor_transition_beta}  shows the capital composition $\{(\empiri{z}_i,c_i)\}_{i=1}^{N}$ for all agents in the stationary state, where $\empiri{z}_i$ is the average number of goods owned by agent $i$. The population of agents separates into two distinct classes: a \textit{class of cash-poor} agents, who own an average number of goods that is very close to the maximum allowed by their wealth, and a \textit{cash-rich class}, where agents have on average the same number of goods.
These two classes are separated by a sharp crossover region. 
The inset of Figure \ref{Fig:Picturesque_RichPoor_transition_beta} shows 
the cash distribution $P_i(\ell/\pi)$ (where $\ell/\pi = c_i/\pi - z$ represents the number of goods they are able to buy) for some representative agents. While cash-poor agents have a cash distribution peaked at $0$, the wealthiest agents have cash in abundance.

These two observations allow us to trace the origin of the arrest in the economy back to the shrinkage of the \textit{cash-rich class} to a vanishingly small fraction of the population, as $\beta \to 1^{+}$. As we'll see in the next section, when $\beta$ is smaller than $1$ the fraction of agents belonging to this class vanishes as $N \to \infty$. In this regime, not only the wealthiest few individuals own a finite fraction of the whole economy's wealth, as observed in Ref. \cite{bouchaud2000wealth}, but they also drain all the financial resources in the economy.

\begin{figure*}
\begin{tabular}{cc}
\includegraphics[width=0.49\textwidth]{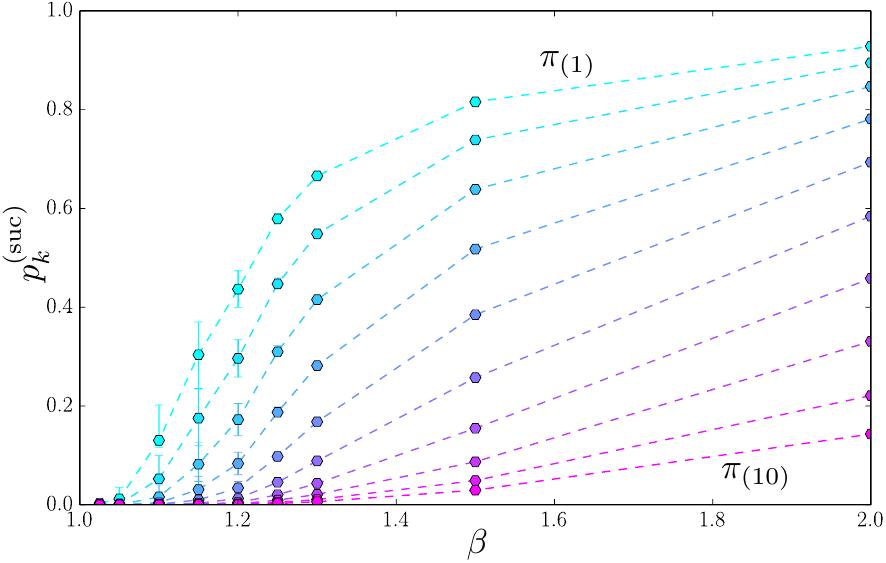}
\includegraphics[width=0.51\textwidth]{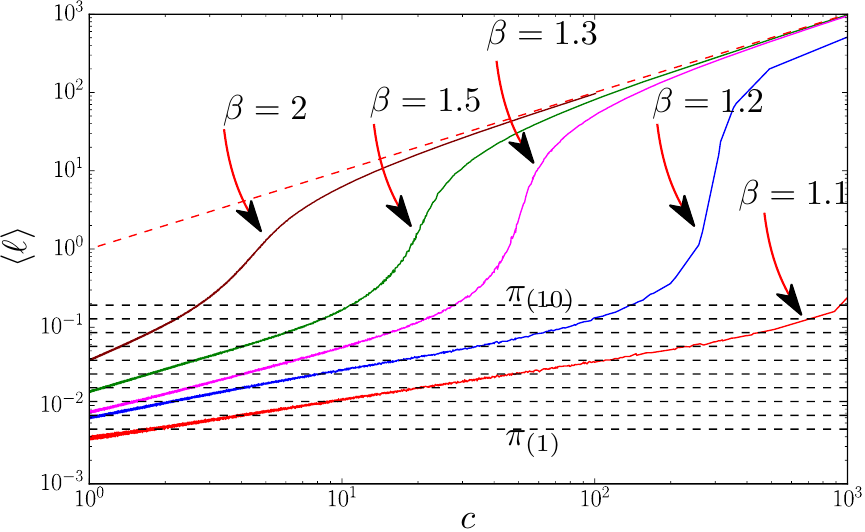}
\end{tabular}
\caption{Left:
 Liquidity of goods $\{p^{(\text{suc})}_{k}\}_{k=1}^K$  as a function of the inequality exponent $\beta$ for a system of $N=10^5$ agents exchanging $K=10$ classes of goods ($\pi_{(k)}=\pi_{(1)}g^{k-1}$ with $g=1.5$, $\pi_{(1)}=0.005$, $M_k\pi_{(k)}=\Pi/K$ and $C/\Pi=1.2$). Note that all success rates $p^{(\text{suc})}_{k}$  vanish when $\beta \to 1^+$.  The curves are ordered from the cheapest (top) to the most expensive (bottom). The markers are the result of numerical simulations, with error bars indicating the minimum and maximum values obtained by averaging over 5 realizations of the wealth allocations (for more details on the simulations see Appendix \ref{app:Detailsonthenumericalmethods}). 
Right: for the same simulations with $K=10$ classes of goods, we plot the time averaged cash $\empiri{\ell_i}$ as a function of wealth $c_i$, from $\beta= 1.1$ to $\beta =2$. The dashed lines indicate the different prices of goods. Agents with $\empiri{\ell_i}$ below the price of a good typically have not enough cash to buy it. 
 Cash is proportional to wealth for large levels of wealth (see the upper straight red dashed line). 
}
\label{Fig:K10_ps_beta}
\end{figure*}

These findings extend to more complex settings. Figure \ref{Fig:K10_ps_beta} illustrates this for an economy with $K=10$ classes of goods (see figure caption for details) and different values of $\beta$. In order to visualise the freezing of the flow of goods we introduce the success rate of transactions for goods belonging to class $k$, denoted as $p^{(\text{suc})}_{k}$.  Figure \ref{Fig:K10_ps_beta} shows that, as expected, for a fixed value of the Pareto exponent $\beta$ the success rate increases as the  goods become cheaper, as they are easier to trade.  Secondly it shows that trades of all classes of goods halt as $\beta$ tends to unity, that is when wealth inequality becomes too large, independently of their price.

The decrease of $p^{(\text{suc})}_{k}$ when inequality increases (i.e. as $\beta$ decreases) is a consequence of the concentration of cash in the hands of the wealthiest agents. This can be observed in the right panel of  Figure \ref{Fig:K10_ps_beta}, which shows the average cash of agents with a given wealth, for different values of $\beta$. The freezing of the economy when $\beta$ decreases occurs because fewer and fewer agents can dispose of enough cash (i.e. have $\ell > \pi_{(k)}$) to buy the different goods (prices $\pi_{(k)}$ correspond to the dashed lines). 

Note finally that $p^{(\text{suc})}_{k}$ quantifies liquidity in terms of goods. In order to have an equivalent measure in terms of cash that can be compared to the velocity of money, we average $\pi_{(k)} p^{(\text{suc})}_{k}$ over all goods
\begin{equation}
\label{def:pavg}
\bar p^{(\text{suc})}=\frac{1}{\Pi}\sum_{k=1}^K M_k \pi_{(k)} p^{(\text{suc})}_k.
\end{equation}
This quantifies the frequency with which a unit of cash changes hand in our model economy, as a result of a successful transaction. It's behaviour as a function of $\beta$ for the same parameters of the economy in Figure \ref{Fig:K10_ps_beta} is shown in the right panel of Figure \ref{Fig:data}.

\section{The analytical approach to the stationary state}
\label{sec:solutionsNumerical_and_analytical}

In order to shed light on the findings described above, in this section we describe how to derive them within an analytic approach. We start by dealing with the simpler case where all the goods in the system have the same price $\pi_m = \pi$, $\forall m$ (i.e. $K=1$). 

A formal approach to this problem consists in writing the complete Master Equation that describes the evolution of the probability $P(z_1,\ldots, z_N)$ to find the economy in a state where each agent $i=1,\ldots,N$ has a definite number $z_i$ of goods. Taking the sum over all values of $z_j$ for $j\neq i$, one can derive the Master Equation for a single agent with wealth $c_i$. The corresponding marginal distribution $P_i(z)$ in the stationary state can be derived from the detailed balance condition
\begin{align}
P_i(z+1)  \frac{z+1}{M}  p^{(\text{suc})}&=  P_i(z) \frac{1}{N} \left(1-\delta_{z, m_{i}}\right) ,\qquad z=0,1,\ldots, m_i
\label{Eq:MasterEq3_1Good_1Guy}
\end{align}
where $m_{i}=\floor{c_i / \pi}$ is the maximum number of goods which agent $i$ can buy with wealth $c_i$ and $p^{(\text{suc})}$ is the probability that a transaction where agent $i$ sells one good (i.e. $z+1\to z$) is successful. Eq. (\ref{Eq:MasterEq3_1Good_1Guy}) says that, in the stationary state, the probability that agent $i$ has $z$ objects and buys a new object is equal to the probability to find agent $i$ with $z+1$ objects, selling successfully one of them. The factor $1-\delta_{z, m_{i}}$ enforces the condition that agent $i$ can afford at most $m_i$ goods and it implies that $P_i(z)=0$ for $z>m_i$. Exchanges are successful if the buyer $j$ does not already have a saturated budget $z_j=m_j$. So the probability $p^{(\text{suc})}$ is also given by
\begin{eqnarray}
p^{(\text{suc})} & = & 1-\frac{1}{N-1}\sum_{j\neq i} P\{z_j=m_j|z_i=z\}\\
 & \cong & 1-\frac{1}{N}\sum_{j} P_j(m_j)\qquad (N,M\gg 1)\label{eq:ps_1Guy}
\end{eqnarray}
where the last relation holds because when $N,M\gg 1$ the dependence on $z$ becomes negligible.
This is important, because it implies that for $N$ large the variables $z_i$ can be considered as independent, i.e. $P(z_1,\ldots, z_N)=\prod_i P_i(z_i)$, and the problem can be reduced to that of computing the marginals $P_i(z_i)$ self-consistently.

The solution of Eq.  (\ref{Eq:MasterEq3_1Good_1Guy}) can be written as a truncated Poissonian with parameter $\lambda = M / (N p^{(\text{suc})}) $
\begin{align}
&P_i(z) = \frac{1}{Z_i} \left[ \frac{\lambda^{z}}{z!}\right] \Theta\left(m_i - z \right)
\label{Eq:ME_solution_1Good}
\end{align}
with $Z_i$ is a normalization factor that can be fixed by $ \sum_{z} P_i(z) = 1$. 
Finally, the value of $p^{(\text{suc})}$ -- or equivalently of $\lambda$ -- can be found self-consistently, by solving Eq. (\ref{eq:ps_1Guy}). 

Notice that the most likely value of $z$ for an agent with $m_i=m$ is given by
\begin{align}
z^{\text{mode}}(m)  \equiv \text{arg}\max_{z}  P(z)=
\begin{cases}
    m, & \text{ if } m \leq \lambda \\
    \lambda, & \text{ if } \lambda \leq m
  \end{cases}.
  \label{Eq:cases_ztyp}
\end{align}
This provides a natural distinction between cash-poor agents -- those with $m\le \lambda$ --  that often cannot afford to buy further objects, and cash-rich ones -- those with $m> \lambda$ --  who typically have enough cash to buy further objects. 

This separation into two classes of agents was already pointed out in Figure \ref{Fig:Picturesque_RichPoor_transition_beta}.
In terms of wealth, the poor are defined as those with $c_i<c^{(1)}$ whereas the rich ones have $c_i>c^{(1)}$, where the threshold wealth is given by $c^{(1)} = \lambda \pi = M \pi/(N p^{(\text{suc})})$. Notice that when $\lambda\gg 1$, a condition that occurs when the economy is nearly frozen ($p^{(\text{suc})}\ll 1$), the distribution $P_i(z)$ is sharply peaked around $z^{\text{mode}}(m)$ so that its average is $\langle z\rangle\simeq z^{\text{mode}}(m)$. Then the separation between the two classes becomes rather sharp, as in Figure \ref{Fig:Picturesque_RichPoor_transition_beta}.

In this regime, we can also derive an estimate of $p^{(\text{suc})}$ in the limit $N\to \infty$, for $\beta>1$. Indeed, we have $P_i(z=m_i)\simeq 1-\frac{m_i}{\lambda}+O(\lambda^{-2})$ for $\lambda\gg m_i$, so a rough estimate of $P_j(m_j)$ is given by 
$P_j(m_j) \simeq \max\{0,1-m_j/\lambda\}$. 
Taking the average over agents, as in Eq. (\ref{eq:ps_1Guy}), and assuming a distribution density of wealth $\rho(c)=\beta c^{-\beta-1}$ for $c\ge 1$ and $\rho(c)=0$ for $c<1$, one finds (see Appendix \ref{app:otherDerivationOfps_using_BalanceInTheMasterEqFashion})
\begin{align}
c^{(1)} &\simeq \left[\beta \lp  1 - \frac{\Pi}{C}\rp \right]^{1/(1 - \beta)}, 
\label{Eq:c1_correct}
\\ 
p^{(\text{suc})} &= \frac{M}{N \lambda} \simeq \frac{\Pi}{C} \frac{\expect{c}}{c^{(1)}}.
\label{Eq:ps_correct}
\end{align}
Here $\expect{c}= \beta/(\beta-1)$ is the expected value of the wealth. Notice that $\expect{c}$ diverges as $\beta \to 1^+$, but also that within this approximation the threshold wealth $c^{(1)}$ diverges much faster, with an essential singularity. 
More precisely, we note that $\Pi/C<1$, so that $\beta ( 1- \Pi/C) \sim (1-\Pi/C)$ is a number smaller than 1 (yet positive).
From Eq. (\ref{Eq:c1_correct}), we have $c^{(1)}\sim (1-\Pi/C)^{-1/(\beta-1)} \to \infty$. 
Therefore the liquidity $p^{(\text{suc})}$ vanishes as $\beta \to 1^+$.

For finite $N$, this approximation breaks down when $\beta$ gets too close to or smaller than one. 
Also, $\expect{c}$ is ill-defined and in Eq. (\ref{Eq:ps_correct}) it should be replaced with $\empiri{c}\equiv 1/N \sum_i c_i $, which strongly fluctuates between realizations and depends on $N$.
An estimate of $p^{(\text{suc})}$ for finite $N$ and $\beta<1$ can be obtained by observing that the wealth $c^{(1)}$  marking the separation between the two classes cannot be larger than the wealth $c_{\max}$ of the wealthiest agent. 
By extreme value theory, the latter is given by $c_{\max}\sim N^{1/\beta}$, with $a>0$. Therefore the solution is characterised by $ c^{(1)}=\pi \lambda\sim c_{\max}\sim N^{1/\beta}$. Furthermore, for $\beta<1$ the average wealth is dominated by the wealthiest few, i.e. $\empiri{c} \sim N^{1/\beta-1}$ and therefore  $p^{(\text{suc})} \sim \empiri{c}/c^{(1)} \sim N^{-1}$.
In other words, in this limit the cash-rich class is composed of a finite number of agents, who hold almost all the cash of the economy.
Figure \ref{Fig:summary_section_one_object} (left) shows that the rough analytical estimate of Eq. (\ref{Eq:ps_correct}) is in good agreement with Monte Carlo simulations.

The analysis carries forward to the general case in which $K$ classes of goods are considered, starting from the full Master Equation for the joint probability of the ownership vectors $\vec z_i=(z_{i,1}\ldots, z_{i,K})$ for all agents $i=1,\ldots,N$. For the same reasons as before, the problem can be reduced to that of computing the marginal distribution $P_i(\vec z_i)$ of a single agent. The main complication is that the maximum number $m_{i,k}$ of goods of class $k$ that agent $i$ can get now depends on how many of the other goods agent $i$ owns, i.e. $m_{i,k}(z^{(k)}_i)=\floor{(c_i-\sum_{k'(\neq k)} z_{i,k'} \pi_{(k')})/\pi_{{k}}}$, where $z^{(k)}_i=\{z_{i,{k'}}\}_{k'(\neq k)}$. The detailed balance condition 
\begin{equation}
\label{Eq:MasterEq3_2Goods_1Guy}
P_i(\vec z+\hat e_k)\frac{z_k+1}{M}p^{(\text{suc})}_{k}=
P_i(\vec z)  \frac{M_k}{M} \frac{1}{N} \left(1-\delta_{z_k,m_{i,k}(z_{(k)})}\right) 
\end{equation}
again yields the stationary state distribution (for $N,M\gg 1$). On the left we have the probability that one of the $z_k+1$ objects of type $k$ of agent $i$ is picked for a successful sale (here $\hat e_k$ is the vector with all zero components and with a $k^{\rm th}$ component equal to one, and $p^{(\text{suc})}_{k}$ is the probability that a sale of an object of type $k$ is successful). This must balance the probability (on the r.h.s.) that agent $i$ is selected as the buyer of an object of type $k$, which requires that agent $i$ has less than $m_{i,k}(z_{(k)})$ objects of type $k$, for the transaction to occur (here $M_k/M$ is the probability that an object of type $k$ is picked at random, and $1/N$ is the probability that agent $i$ is selected as the buyer).
It can easily be checked that the solution to this set of equations is given by a product of Poisson laws with parameters $\lambda_k=M_k/(N p^{(\text{suc})}_{k})$, with the constraint Eq. (\ref{eq:1}), 
\begin{align}
&P_i(z_1,..., z_{K}) = \frac{1}{Z_i} \left[ \prod_{k=1}^{K}\frac{\lambda^{z_k}_k}{z_k!}\right] \Theta\left(c_i - \sum_k^{K} z_k \pi_{(k)}\right)\,,
\label{Eq:ME_solution_2Goods}
\end{align}
with $Z_i$ a normalization factor obeying $ \sum_{z_1}...  \sum_{z_{K}}  P_i(z_1, ..., z_{K}) = 1$. Here the $p_k^{(\text{suc})}$ corresponds to the acceptance rates of transactions of goods of class $k$ and are given by
\begin{align}
&p^{(\text{suc})}_{k} = 1 - \frac{1}{N}\sum_{i=1}^N P\left\{z_{i,k}=m_{i,k}(z_i^{(k)})\right\}\,
\label{eq:ps_Kobjs}
\end{align}
As in the case with $K = 1$, the values of the $p^{(\text{suc})}_{k}$ need to be found self-consistently, which can be complicated when $K$ and $M$ are large. 

When  the total number of objects per agent is large for any class $k$, we expect that $\lambda_1, ..., \lambda_K \gg 1$, and then the values of $z_{i,k}$ are close to their expected values. This implies that the population of agents splits into $K$ classes, where agents with wealth $c_i\in [c^{(k-1)},c^{(k)}]$ have their budget saturated with goods of class $k'\le k$ and cannot afford more expensive objects (here $c^{(k)}=\lambda_k\pi_{(k)}$, $k=1,\ldots, K$ and $c^{(0)}=c_{\min}$). An estimate for the thresholds $c^{(k)}$ can be derived following the same arguments as for 
$K=1$,
 by observing that when analysing the dynamics of goods of type $k$, all agents in class $k'<k$ are effectively frozen and can be neglected. 
Combining this with the conservation of the total number of objects of each kind, we obtain a recurrence relation for $c^{(k)}$. We refer the interested reader to the Appendix \ref{app:recurrence} for details on the derivation, and report here the result in the case of goods with $\pi_{(k)}=\pi_{(1)} g^{k-1}$, $g> 1$ large enough, with $\beta>1$ and in the limit $N\to\infty$:
\begin{align}
c^{(k)} &\simeq \left[ \beta^k  - \lp \frac{\beta - \beta^{k+1}}{1-\beta} \rp \frac{\Pi}{K C} \right] ^{\frac{1}{1-\beta}},
 \label{Eq:ps_and_ck_generalCaseAnyMathcalM1}
\\
p_k^{(\text{suc})} &= \frac{M_k}{N \lambda_k} \simeq \frac{\Pi}{K C} \frac{\expect{c}}{c^{(k)}}.
 \label{Eq:ps_and_ck_generalCaseAnyMathcalM2}
\end{align}

In the limit  $\beta \to 1^+$ of large inequality, close inspection\footnote{Note that the term in square brackets is smaller than one, when $\beta \to 1^+$.} of Eq. (\ref{Eq:ps_and_ck_generalCaseAnyMathcalM1}) shows that $ c^{(k)}  \to \infty, \forall k$, 
which implies that all agents become cash-starved except for the wealthiest few. 
Since $p^{(\text{suc})}_{k}\sim \expect{c} /c^{(k)}$, this implies that all markets freeze: $p^{(\text{suc})}_{k}\to 0 , \forall k$.  
The arrest of the flow of goods appears to be  extremely robust against all choices of the parameter $\pi_{(k)}$, as $p^{(\text{suc})}_{1}$ is an upper bound for the other success rates of transactions $p^{(\text{suc})}_{k}$. These conclusions are fully consistent with the results of extensive numerical simulations (see Figure \ref{Fig:summary_section_one_object} in appendix \ref{app:DerivationOfps}).

\section{Summary and conclusions}
\label{sec:con}
In this paper we have introduced a zero-intelligence trading dynamics in which agents have a Pareto distributed  wealth and randomly trade goods with different prices. We have shown that this dynamics leads to a uniform distribution in the space of the allocations that are compatible with the budget constraints.  We have also shown that when the inequality in the distribution of wealth increases, the economy converges to an equilibrium where typically (i.e. with probability very close to one) the less wealthy agents have less and less cash available, as their budget becomes saturated by objects of the cheapest type. At the same time this class of cash-starved agents takes up a larger and larger fraction of the economy, thereby leading to a complete halt of the economy when the distribution of wealth becomes so broad that its expected average diverges (i.e. when $\beta \to 1^{+}$). In these cases, a finite number of the wealthiest agents own almost all the cash of the economy. 

The model presented in this paper is intentionally simple, so as to highlight a simple, robust and quantifiable link between inequality and liquidity. In particular, the model neglects important aspects such as {\em i)} agents' incentives and preferential trading, {\em ii)} endogenous price dynamics and {\em iii)} credit. 
It is worth discussing each of these issues in order to address whether the inclusion of some of these factors would revert our finding that inequality and liquidity are negatively related. 

First, our model assumes that all exchanges that are compatible with budget constraints will take place, but in more realistic setting only exchanges that increase each party's utility should take place. Yet if the economy freezes in the case where agents would accept all exchanges that are compatible with their budget, it should also freeze when only a subset of these exchanges are feasible. Also the model assumes that all agents trade with the same frequency whereas one might expect that rich agents trade more frequently than poorer ones. Could liquidity be restored if trading patterns exhibit some level of homophily, with rich people trading more often and preferentially with rich people? 

First we note that both these effects are already present in our simple setting. Agents with higher wealth are selected more frequently as sellers as they own a larger share of the objects. In spite of the fact that buyers are chosen at random, successful trades occur more frequently when the buyer is wealthy. 
So, in the trades actually observed the wealthier do trade more frequently than the less wealthy, and preferentially with other wealthy agents. 
Furthermore, if agents are allowed to trade only with agents having a similar wealth (e.g. with the $q$ agents immediately wealthier or less wealthy) it is easy to show that detailed balance still holds with the same uniform distribution on allocations. As long as all the states are accessible, the stationary probability distribution remains the same\footnote{
The dynamics changes and thus $p_k^{(\text{suc})}$ changes, in particular for goods more expensive than $\pi_{(1)}$, the seller is typically cash-rich and thus its neighbours are too. This can induce to have a liquidity of expensive goods higher than that of cheaper ones. However in the limit $\beta \to 1^{+}$, it is still true that cash concentrates in the hands of a vanishing fraction of agents, and there is still a freeze of the economy.}. Therefore, our conclusions are robust with respect to a wide range of changes in our basic setting that would account for more realistic trading patterns. 

Secondly, it is reasonable to expect that prices will adjust -- i.e. deflate -- as a result of a diminished demand caused by the lack of liquidity. 
Within our model, the inclusion of price adjustment, occurring on a slower time-scale than trading activity, would reduce the ratio $\Pi/C$ (between total value of goods and total wealth), but it would also change the wealth distribution. 
Since the freezing phase transition occurs irrespective of the ratio $\Pi/C$, the first effect, though it might alleviate the problem, would not change our main conclusion. The second would make it more compelling, because cash would not depreciate as prices do, so deflation would leave 
wealthy 	
agents -- who hold most of the cash -- even richer compared to the cash deprived agents, that would suffer the most from deflation. So while price adjustment apparently increases liquidity, this may promote further inequality, that would curtail liquidity further. 

Finally, can the liquidity freeze be avoided by allowing agents to borrow? Access to credit, we believe, will hardly improve the situation\footnote{Allowing agents to borrow using goods as collaterals is equivalent to doubling the
wealth of cash-starved agents, provided that any good can be used only once as a collateral, and that goods bought with credit cannot themselves be used as collaterals.  This would at most blur the crossover between cash-rich agents and cash-starved ones, as intermediate agents would sometimes use credit. This does not change our main conclusion that inequality and liquidity are inversely related and that the economy would halt when $\beta \to 1^{+}$.}, in line with the results of Ref. \cite{Yakovenko2009Review} and for similar reasons. Credit may mitigate illiquidity in the short term, but cash deprived agents should borrow from wealthier ones. With positive interest rates, this would make inequality even larger in the long run. So credit is likely to make things worse, in line with the arguments\footnote{Piketty \cite{Piketty2014} observes that when the rate of return on capital exceeds the growth rate of the economy (which is zero in our setting), wealth concentrates more in the hands of the rich.} in \cite{Piketty2014}.

Therefore, even though the model presented here can be enriched in many ways, we don't see a way in which the relation between inequality and liquidity could be reversed. 

Corroborating the present model with empirical data goes beyond the scope of the present paper, yet we remark that our findings are consistent with the recent economic trends, as shown in Figure \ref{Fig:data}.  For example,  it is worth observing that, alongside with increasing levels of inequality, trade 
has slowed down after the 2008 crisis\footnote{The {\em U.S. Trade Overview, 2013} of the International Trade Administration observes that ``Historically, exports have grown as a share of U.S. GDP. However, in 2013 exports contributed to 13.5\% of U.S. GDP, a slight drop from 2012'" (see {\tt http://trade.gov/mas/ian/tradestatistics/index.asp{\#P}11}). A similar slowing down can be observed at the global level, in the UNCTAD {\em Trade and Development Report, 2015}, page 7 (see {\tt http://unctad.org/en/pages/PublicationWebflyer.aspx?publicationid=1358}).}. More generally, avoiding deflation -or promoting inflation- has been a major target of monetary policies after 2008, which one could take as an indirect evidence of the slowing down of the economy.  Furthermore, the fact that inequality hampers liquidity and hence promotes demand for credit suggests that the boom in credit market before 2008 and the increasing levels of inequality might not have been a coincidence. 

An interesting side note is that the concentration of capital in the top agents goes hand in hand with a flow of cash to the top.  Indeed, in our model an injection of extra capital in the lower part of the wealth pyramid --the so-called {\em helicopter money} policy-- is necessarily followed by a flow of this extra cash to the top, via many intermediate agents, thus generating many transactions on the way. This \textit{trickle up} dynamics should be contrasted with the usual idea of the \textit{trickle down} policy, which advocates injections of money to the top in order to boost investment. 
In this respect, it is tempting to relate our findings to the recent debate on Quantitative Easing measures, and in particular to the proposal that the (European) central bank should finance households (or small businesses) rather than financial institutions in order to stimulate the economy and raise inflation \cite{QEvox,QEft}. 
Clearly, our results support the helicopter money policy, because injecting cash at the top does not disengages the economy from a liquidity stall. 


Extending our minimal model to take into account the endogenous dynamics of the wealth distribution and of prices, accounting for investment and credit, is an interesting avenue of future research, for which the present work sets the stage. In particular, this could shed light on understanding the conditions under which the positive feedback between returns on investment and inequality, that lies at the very core of the dynamics which has produced ever increasing levels of inequality according to \cite{Piketty2001,Piketty2014,SaezZucman2016}, sets in.

\section{Acknowledgments}
JPJ thanks FAPESP process number 2014/16045-6 for the financial support and the Abdus Salam ICTP for the hospitality. IPC thanks the Abdus Salam ICTP for the hospitality and also thanks Javier Toledo Mar\'in for the work done during some stages of the research presented here. MM thanks Arnab Chatterjee for discussions at an earlier stage of this project and Davide Fiaschi for interesting discussions and comments.

 

\newpage
\appendix
\section{Appendix}

\subsection{About the Rules Providing Detailed Balance}
\label{app:rules_enumeration}

The detailed balance condition is a useful criterium to find the stationary state in stochastic processes. Given a dynamics formulated in terms of the transition rates $W(\mathcal{A}_i,\mathcal{A}_j)$ between configurations $\mathcal{A}_i$ and $\mathcal{A}_j$, 
If one can find a measure $P(\mathcal{A}_i)\ge 0 $ over configurations that satisfies the detailed balance condition
\begin{align}
 \forall i,j, \qquad 
W(\mathcal{A}_i,\mathcal{A}_j) P(\mathcal{A}_i) = W(\mathcal{A}_j,\mathcal{A}_i)P(\mathcal{A}_j),
\end{align}
and if the system is ergodic\footnote{Meaning that for each pair of configurations $\mathcal{A}_i$ and $\mathcal{A}_j$ there is a path of a finite number of intermediate configurations $\mathcal{A}_{i_k}$ with non-zero rate $W(\mathcal{A}_{i_k},\mathcal{A}_{i_{k+1}})$,},
then $P(\mathcal{A}_i) $ is the unique stationary distribution. 
The detailed balance property corresponds to a local balance of the flux between any pair of configurations.

The simplest way to have detailed balance property is to use symmetrical transfer rates: $W(\mathcal{A}_i,\mathcal{A}_j)=W(\mathcal{A}_j,\mathcal{A}_i)$.
In that case, one automatically gets a uniform distribution over the space of configurations: $P(\mathcal{A})=const, \forall \mathcal{A} $. 
The flux $W(\mathcal{A}_1,\mathcal{A}_2) P(\mathcal{A}_1)$ is then also uniform.
It is clear that the dynamics defined in this paper has this property, because for any two configurations that differs by the ownership of one object, the rate of the process linking them is equal to $1/(NM)$ in both directions.

What about the rules providing detailed balance, but without symmetry of the rates?  In that case, one would need to explicitly find the probability density over the configurations. 
Since the resulting density would be non-uniform, it would be more difficult to link dynamical observables (rate of money transfer, etc.) to static variables (number of neighbouring configuration to a given configuration). We do not explore these cases.

What are the rules that give symmetrical transfer rates? 
We define time such that at each time step, a single object is picked and there is a single attempt at selling it to another agent (not the current owner).
In this paper, we consider the simplest case where objects are picked independently of their price\footnote{One could pick an object with a rate proportional to its value. This kind of choice would still give the same phase space and thus the same probability distribution over microstates, but the dynamics could become very different in terms of the speed of transactions, in particular it could fluctuate much more.}. 
This still leaves us several choices. 
There are $N-1$ rules which yield symmeteric rates (and thus respect detailed balance). The generic case is the following, with $2\leq n \leq N$:
\begin{itemize}
\item
rule $\# n$: The integer $n$ is fixed. Select $n$ distinct agents at random. Select one object among the set of all the objects they (collectively) own. This object will be sold (if possible) by the owner to a randomly selected agent among the $n-1$ remaining agents in the set of selected agents.
\end{itemize}

This generic rule is a bit cryptic, but has two particular cases that are clearer:
\begin{itemize}
\item rule $\# 2$: Select two distinct agents at random. Select one object among the set of all the objects they (collectively) own. This object is sold (if possible) by the owner to the other agent.
\item rule $\# N$: Pick an object at random. The owner is then the seller. Select a buyer at random amongst the $N-1$ remaining agents.
\end{itemize}
Note that the rule $\#n = 1$ does not make sense, so that there are indeed $N-1$ different rules.
In this paper, we always use the rule $\# N$, i.e. simply pick the object at random.
As all these rules produce an ergodic dynamics, and since the probability distribution of configurations is the same for all rules (it is $P(\mathcal{A})=const.$), it does not matter at all which of these dynamical rules we picked.

We do not claim that these rules are the only ones providing symmetric rates, or even that this property is necessary for interesting dynamics.
We simply point out that the detail of the choice of the dynamical rule is not crucial, as long as we as we have a simple zero-intelligent dynamics which generates symmetric transition rates.
We now give a few examples of rules that are either identical to those above, or do not yield symmetrical probabilities of transfer:
\begin{itemize}
\item Select one seller among the N agents by using weighted probabilities: each agent $i$ is affected a weight $n_o$ where $n_o$ is the number of objects owned by agent $i$. Pick an object from this agent at random. It will be sold to an agent picked uniformly among the $N-1$ remaining agents. This rule is actually exactly the rule $\# N$.
\item The following rule does not yield symmetrical probabilities of transfer: select a buyer at random. Select any object not already owned by him at random. The owner of that object sells it to the buyer. The problem is that if the buyer has more objects than the seller, the inverse transaction has smaller probability to occur than the direct one.
\item The following rule very clearly breaks symmetry: select one seller at random. Select an object in his set at random. Select a buyer among the $N-1$ remaining agents at random. Naively, this rule could be named $\# 1$, but it clearly breaks symmetry. (The transaction from a hoarder to an agent with few objects has much smaller probability to occur than the inverse transaction.)
\end{itemize}

\subsection{Computation of  $p^{(\text{suc})}_k$ in the large $\lambda$ limit}
\label{app:DerivationOfps}

\subsubsection{Derivation of $p^{(\text{suc})}$ and $c^{(1)}$ in the large $\lambda$ limit for 1 type of good.}
\label{app:otherDerivationOfps_using_BalanceInTheMasterEqFashion}

As discussed in the main text, we can compute $p^{(\text{suc})}$ using
\begin{equation}
\label{eq:app_ps_analytic0}
p^{(\text{suc})} = 1 - \frac{1}{N} \sum_{i=1}^{N} P_i(z = m_i) 
\end{equation}
approximating the probability to be on a threshold $P_i(z=m_i)$ by
\begin{align}
P_i(z=m_i) = 
\begin{cases}
\left( 1 - \frac{m_i}{\lambda} \right) & \text{ for } m_i \ll \lambda \\
0 & \text{ for } m_i > \lambda
\end{cases}.
\end{align}
The first case can be understood by noting that
\begin{equation}
P_i(z=m_i) = \frac{\lambda^{m_i} \frac{1}{m_i!}}{\sum_{x=0}^{m_i} \lambda^x \frac{1}{x!}} = \frac{1}{1 + \frac{m_i}{\lambda} + \frac{m_i (m_i -1)}{\lambda^2} + \ldots} \simeq  \left( 1 - \frac{m_i}{\lambda} \right),
\end{equation}
where the approximation is valid in the limit $m_i \ll \lambda$. 
Assuming this approximation to be valid in all the range $m_i < \lambda$ is clearly a bad assumption for all agents with $m_i$ close to $\lambda$. 
However the wealth is power law distributed, and so the weight of agents with $m_i \sim \lambda$ is negligible in the sum over all agents,  Eq.~(\ref{eq:app_ps_analytic0}). 
The accuracy of this approximation increases when the exponent of the power law $\beta$ decreases. 

Then $p^{(\text{suc})}$ can be computed using
\begin{equation}
\label{eq:app_ps_analytic1}
p^{(\text{suc})} = 1 - \frac{1}{N} \sum_{i=1}^{N} P_i(z = m_i) \simeq 1 - \int_1^{c^{(1)} = \lambda \pi} dc\, \beta c^{-\beta - 1} \left( 1 - \frac{c}{\lambda \pi} \right) .
\end{equation}
This is an implicit expression for $p^{(\text{suc})}$, since it appears on the l.h.s. of the equation and also on the r.h.s. (because $\lambda = \frac{M}{N p^{(\text{suc})}}$). 

When $\beta > 1$ this expression can be expressed to be realization-independent, using
\begin{equation}
p^{(\text{suc})} = \frac{M}{N \lambda} = \frac{\Pi}{C} \frac{\expect{c}}{c^{(1)}},
\end{equation} 
where $\expect{c} = \beta/ (\beta - 1)$ is the expected value of the wealth per agent. We also use the fact that we fill in the system a number $M$ of goods in such a way to have a fixed ratio $\Pi / C$. Performing the integral on the r.h.s of Eq.~(\ref{eq:app_ps_analytic1}) gives an equation for $c^{(1)}$:
\begin{equation}
\frac{\Pi}{C} \frac{\expect{c}}{c^{(1)}} = {c^{(1)}}^{-\beta} \left( \frac{1}{1-\beta} \right) - \frac{\beta}{1 - \beta } \frac{1}{c^{(1)}},
\end{equation}
that simplifies into:
\begin{equation}
c^{(1)} = \left[\beta \lp  1 - \frac{\Pi}{C}\rp \right]^{1/(1 - \beta)}.
\end{equation}

\subsubsection{Derivation of $p_k^{(\text{suc})}$ and $c^{(k)}$ in the large $\lambda$ limit for several types of good.}
\label{app:recurrence}

An analytic derivation for the $p_k^{(\text{suc})}$ and $c^{(k)}$ can be obtained also for the cases of several goods, but only in the limit in which prices are well separated (i.e. $\pi_{(k+1)} \gg \pi_{(k)}$) and the total values of good of any class is approximately constant (we use $M_k\pi_{(k)}  = \Pi / K = {\rm const}$). 
In this limit we expect to find a sharp separation of the population of agents into classes. 
This is because $M_1\gg M_2\gg \ldots \gg M_K$ implies that the market is flooded with objects of the class $1$, which constantly change hands and essentially follow the laws found in the single type of object case. 
On top of this dense gas of objects of class $1$, we can consider objects of class $2$ as a perturbation (they are picked $M_2/M_1$ times less often!).
On the time scale of the dynamics of objects of type $2$, the distribution of cash is such that all agents with a wealth less than $c^{(1)}=\pi_{(1)}\lambda_1$ have their budget saturated by objects of type $1$ and typically do not have enough cash to buy objects of type $2$ nor more expensive ones. 
Likewise, there is a class of agents with $c^{(1)}<c_i\le c^{(2)}$ that will manage to afford goods of types $1$ and $2$, but will hardly ever hold goods more expensive that $\pi_{(2)}$. 

In brief, the economy is segmented into $K$ classes, with class $k$ composed of all agents with  $c_i\in [c^{(k)},c^{(k+1)})$ who can afford objects of class up to $k$, but who are excluded from markets for more expensive goods, because they rarely have enough cash to buy goods more expensive than $\pi_{(k)}$. This structure into classes can be read off from Figure \ref{Fig:K10_ps_beta}, where we present the average cash of agents, given their cash in a specific case (see caption). The horizontal lines denote the prices $\pi_{(k)}$ of the different objects, and the intersections with the horizontal lines define the thresholds $c^{(k)}$. Agents that have $c_i$ just above $c^{(k)}$ are cash-filled in terms of object of class $k$, but are cash-starved in terms of objects $\pi_{(k')}, k'>k$. 

The liquidities $p^{(\text{suc})}$ can be given by the following expression
\begin{equation}
p^{(\text{suc})}_{k} = 1 - \frac{1}{N}\sum_{i=1}^N P\left\{z_{i,k}=m_{i,k}(z_i^{(k)})\right\} = 1 - \frac{1}{N}\sum_{i=1}^N P_i(\text{not accepting good type k}) 
\end{equation}
According to the previous discussion of segmentation of the system into $K$ classes, and using the same approximation for this threshold probability discussed in the case of 1 type of good, we assume
\begin{align}
P_i(\text{not accepting good type k}) = 
\begin{cases}
1 & \text{ for } m_i < \lambda_{k-1} \\
\left( 1 - \frac{m_i}{\lambda_k} \right) & \text{ for } \lambda_{k-1} < m_i < \lambda_k \\
0 & \text{ for } m_i > \lambda_k
\end{cases},
\end{align}
Then
\begin{equation}
p^{(\text{suc})}_{k} \simeq 1 - \int_{1}^{c^{(k-1)}} dc\, \beta c^{-\beta - 1} - \int_{c^{(k-1)}}^{c^{(k)}} dc\, \beta c^{-\beta - 1} \left( 1 - \frac{c}{c^{(k)}} \right) 
\end{equation}
In this case now we have
\begin{equation}
p_k^{(\text{suc})} = \frac{M_k}{N \lambda_k} = \frac{\Pi}{K C} \frac{\expect{c}}{c^{(k)}}
\end{equation}
With similar calculations to the ones showed for the previous case, one can easily get to the recurrence relation:
\begin{equation}
c^{(k)} = \left[ \beta \lp c^{(k-1)} \rp^{1-\beta} - \beta \frac{\Pi}{K C} \right] ^{\frac{1}{1-\beta}}.
\end{equation}
Iterating, we explicit this into: 
\begin{equation}
c^{(k)} = \left[ \beta^k  - \lp \frac{\beta - \beta^{k+1}}{1-\beta} \rp \frac{\Pi}{K C} \right] ^{\frac{1}{1-\beta}},
\end{equation}

\begin{figure}
\includegraphics[width=0.5\textwidth]{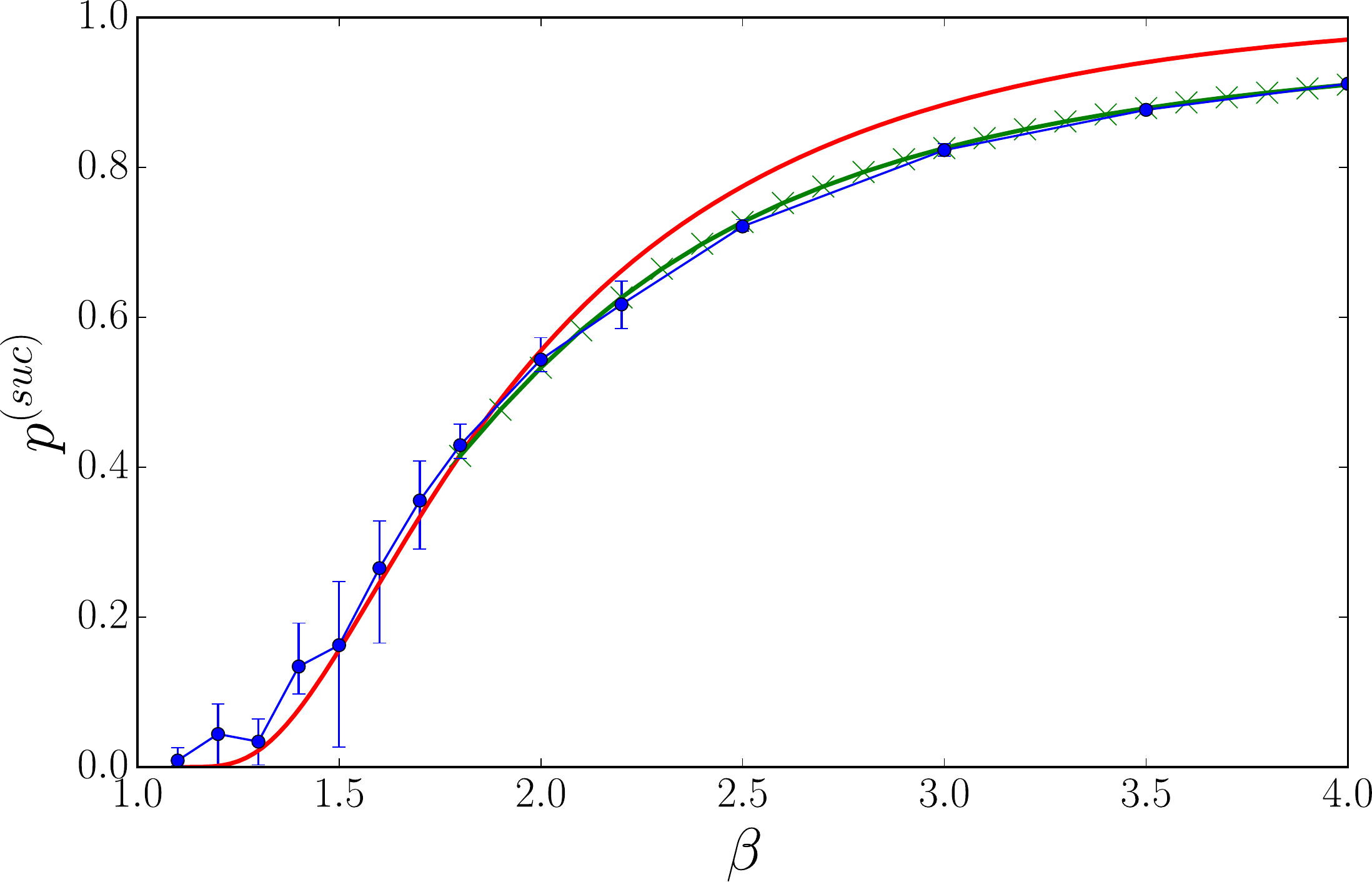}
\includegraphics[width=0.5\textwidth]{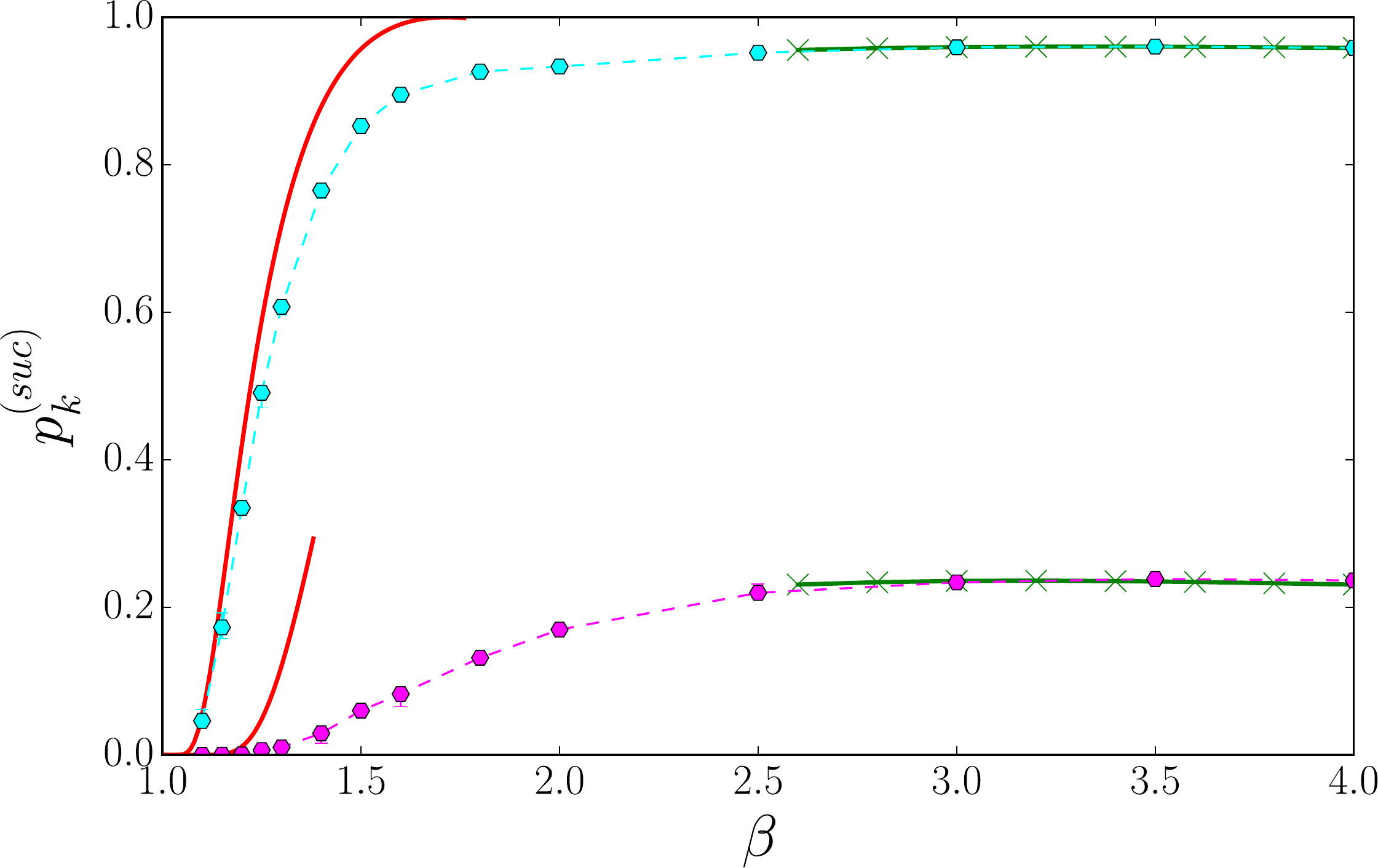}
\caption{Success probability of transaction $p^{(\text{suc})}_k$  as a function of the Pareto exponent $\beta$. Comparison between numerical simulations and analytical estimates for one class of goods (left panel) and two classes of goods (right panel). The blue solid circles are the result  of Monte Carlo simulations performed for $N=10^5$ agents and averaged over 5 realizations. Here the error bars indicate the min and max value of $p^{(\text{suc})}_k$ over all realizations (we used the ``adjusted Pareto'' law for the right panel, see Appendix \ref{app:Adjusted_Pareto_Capital_Distribution}). The red lines are the analytic estimates  according to Eq. (\ref{Eq:ps_correct}) and Eq. (\ref{Eq:ps_and_ck_generalCaseAnyMathcalM1}) for left and right panels, respectively. The green crossed  lines correspond to numerically  (see Appendix \ref{app:iterativeMethodDetailExplained}) solving the analytical solution  \eqref{Eq:ME_solution_2Goods} for a population composed of $N=64$ (kind of) agents.
}
\label{Fig:summary_section_one_object}
\end{figure}

A comparison between the analytical estimate and numerical simulations, presented in Figure \ref{Fig:summary_section_one_object}, shows that this approximation 
provides an accurate description of the collective behaviour of the model. 

\begin{figure}
\centering
\includegraphics[width=8cm]{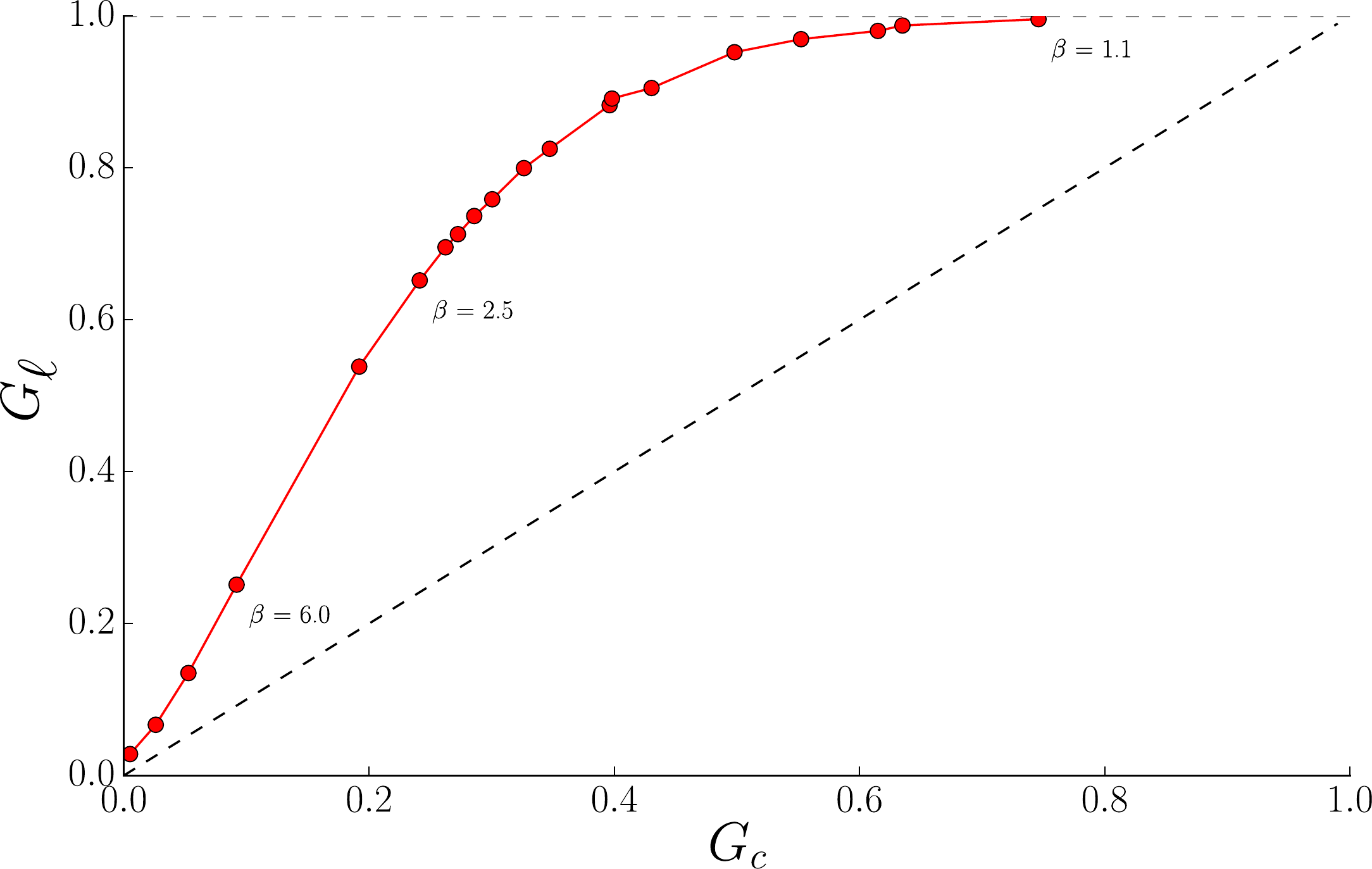}
\caption{Gini coefficient $G_\ell$ of the cash distribution (liquid capital) in the stationary state of the model as a function of the Gini $G_c$ of the wealth distribution. The dashed line indicates proportionality between cash and wealth, in which case the inequality in both is exactly the same.
The wealth follows a Pareto distribution with exponent $\beta$ that tunes the degree of inequality (the higher is $\beta$, the more egalitarian the distribution).} 
\label{Fig:Gini_Intro}
\end{figure}

See also in Fig. \ref{Fig:Gini_Intro} how the liquidity over-concentrates (with respect to capital concentration).
There, we compare the liquid and capital concentrations, measured via their Gini coefficients, for various values of $\beta$ in the system of Fig. \ref{Fig:K10_ps_beta} ($K=10, g=1.5, \pi_{(1)}=0.001, C/\Pi=1.2$).
In particular, note that the limit $\beta\to 1^+$ is singular, as $G_\ell$ reaches one around $\beta=1.1$, with smaller $\beta$ yielding also $G_\ell\approx 1$.
This is an alternative way to see how the concentration of capital generates an over-concentration of liquidities.

\subsection{Details on the numerical methods}

\subsubsection{Monte-Carlo Simulations}
\label{app:Detailsonthenumericalmethods}
\label{app:Adjusted_Pareto_Capital_Distribution}

We perform our Monte Carlo simulations of the trading market for $N=10^5$ agents. 
Prices generally start from $\pi_{(1)}$ and increase by a factor $g$ between each good class.
The minimal wealth is $c_{min}=1$.
The ratio $C/\Pi$ is fixed as indicated in captions, and most importantly is kept constant between different realizations. 
As the total wealth fluctuates, so does the total number of goods.

There are no peculiar difficulties with the numerical method (apart from the large fluctuations in the average wealth, addressed below).
The only thing one has to be careful with is to ensure that the stationary state has been reached, i.e. that all observables have a stationary value, an indication that the (peculiar) initial condition has been completely forgotten. 
The codes for this Monte Carlo simulation are available online \cite{CodeLandes2015}.

\paragraph{Adjusted Pareto Wealth Distribution}

For $K \geq 2$ we have predictions for the $\beta\sim 1$ regime, in which the average wealth is particularly fluctuating from realization to realization.
Because the value of $\expect{c}$ controls the number $M$ of objects introduced in the market, this in turns produces large fluctuations in the values of the $p^{(\text{suc})}_k$ which can make it difficult to have robust results.

\begin{figure}
\centering
\includegraphics[width=0.5\textwidth]{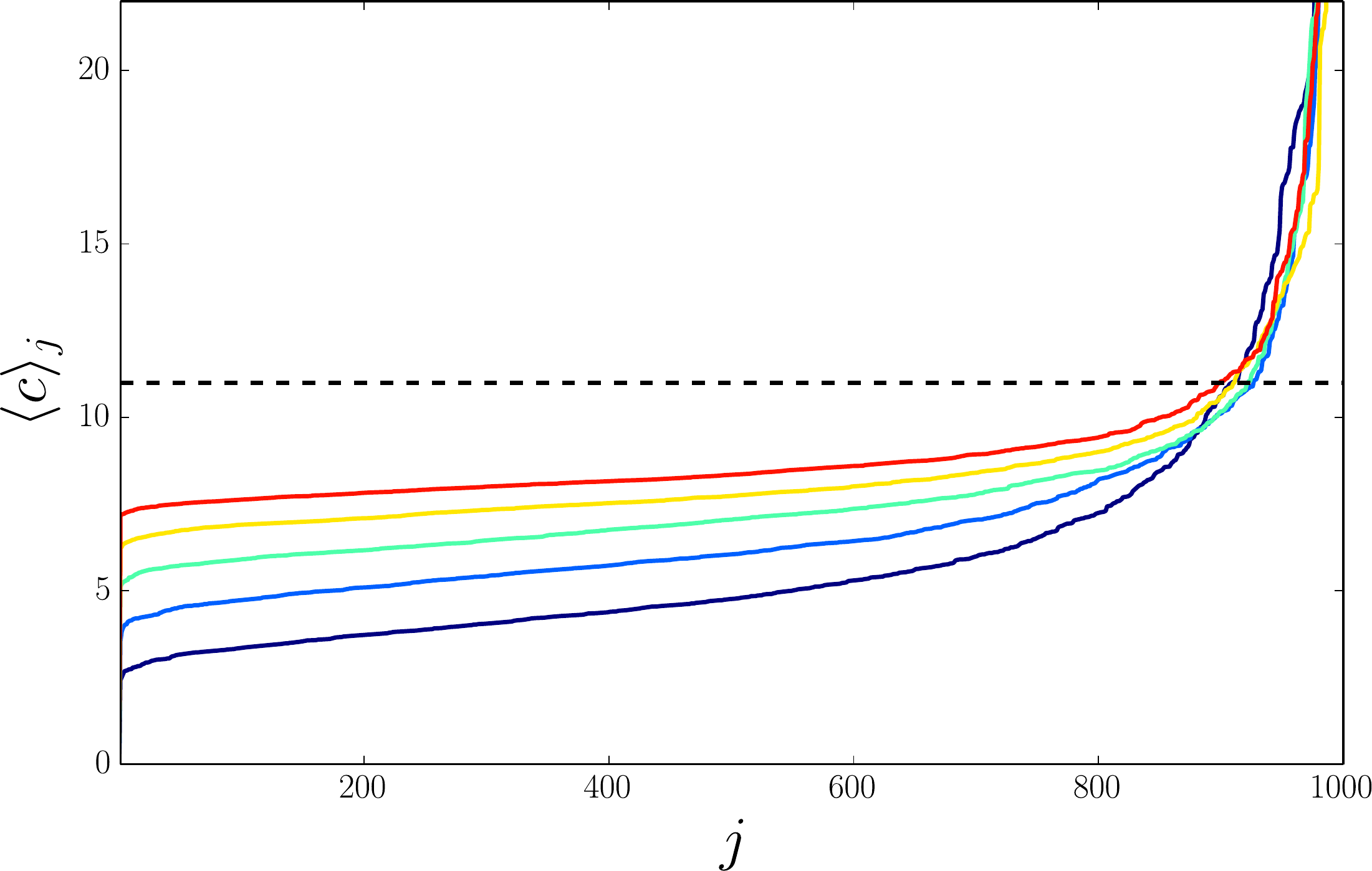}
\caption{
Distribution of the average $\empiri{c}_j$ of power laws depending on their sample size $N$ (from top to bottom, $N=10^6, 10^5, 10^4, 10^3, 10^2$) for 1000 realizations each ($j=1, ... , 1000$), using an exponent $\beta=1.1$.
The dashed line indicates the expectation value $\expect{c} = c_{\min} \beta/(\beta-1)$.
We see that even for huge samples, the typical  $\empiri{c}_j$'s are significantly smaller than the expected  $\expect{c}$.
\label{Fig:nonConvergenceLargeN}
}
\end{figure}

More importantly, the typical value of the (empirical) average wealth $\empiri{c}$ is usually quite different from its expectation value $\expect{c}$.
This effect is well known and well documented for power laws, but we present a concrete example of it in Figure \ref{Fig:nonConvergenceLargeN} to emphasize its intensity.

For the sample size that is typically manageable in our simulations, i.e. $N=10^5$, the typical value for the average value of the wealth (using e.g. $\beta=1.1$) is of the order of the half of its expected value: $\empiri{c}\approx 7\approx \expect{c}/2$.
This indicates that $N=10^5$ is (by far) an insufficient size to correctly sample a power-law with exponent $\beta=1.1$.

To circumvent this problem, we introduce the ``adjusted'' Pareto distribution. 
The idea is to draw numbers from a power law distribution as usual, and then to adjust the value of a few of them so that the empirical average matches the expected one.
The algorithm is the following:
Start from a true random Pareto distribution.
\begin{itemize}
\item if $\empiri{c} < \expect{c}$, we select an agent at random and increase its wealth until we have exactly  $\empiri{c} = \expect{c}$.
\item if $\empiri{c} > \expect{c}$, we select the wealthiest agent and decrease its wealth until we have exactly  $\empiri{c} = \expect{c}$, or until its wealth becomes $c_{min}$. If we reach the latter case (it is quite unlikely), then we perform the same operation on the second-wealthiest agent, and so on until  $\empiri{c} = \expect{c}$.
\end{itemize}
As can be seen in Figure \ref{Fig:nonConvergenceLargeN}, the most common case is the first one. 
The corresponding adjustment is equivalent to re-drawing the wealth of a single agent until it is such that  $\empiri{c} = \expect{c}$. This is a weak deviation from a true Pareto distribution.
The second case is more rare, and mostly consists also in a correction on the wealth of a single agent.

\begin{figure}[htbp]
\centering
\includegraphics[width=\textwidth]{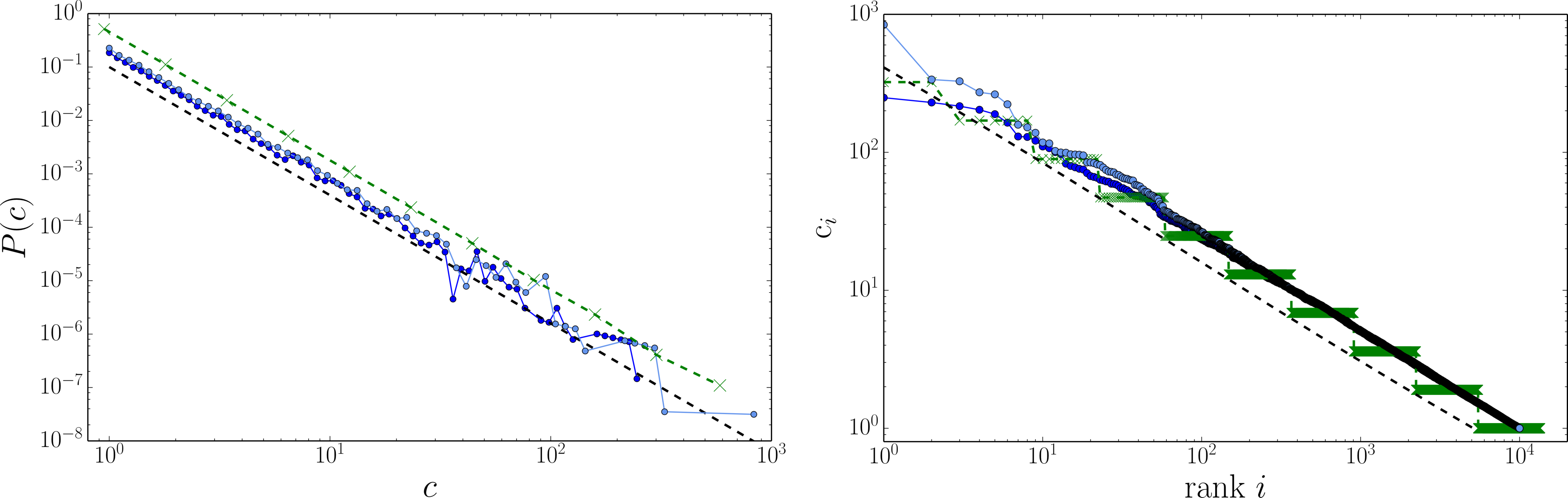}
\caption{
Different instances and representation of power-law distributed wealth (or ``Pareto distribution'').
Blue and pale blue circles are two realisations for $N=10^4$ agents, green crosses are an example of staircase-like distribution (a useful approximation of a Pareto law that we use elsewhere) and the black dashed line is the law itself (blue dots converge to it in the $N\to \infty$ limit).
Left: Probability distribution, with shifts up and down for clarity (i.e., it is not normalized) 
Right: wealth $c_i$ of each agent, sorted by the rank $i$. Note that the wealth of the wealthiest agents (low rank) fluctuates a lot from realization to realization.} 
\label{Fig:Various_Capital_Distro}
\end{figure}

This change in the wealth distribution is very efficient at reducing the variability between different realizations of the same $\beta$ value. 
Furthermore it ensures that we can compare our numerical results at finite $N$ with the predictions that implicitly assume $N=\infty$, since we now have $\empiri{c} = \expect{c}$.
It is quite crucial to use this ``adjusted'' Pareto law for the small $\beta$'s (i.e. for $\beta \leq 1.3$).
See Figure \ref{Fig:Various_Capital_Distro} to have an idea of what this modified distribution means: the only changes in the two sample shown would be in the values of the wealthiest agent.

\subsubsection{Algorithm computing self-consistent solution $p^{(\text{suc})}, Z(c_i)$}
\label{app:iterativeMethodDetailExplained}

\label{subsec:staircases}

Here we describe the algorithm used to converge to a self-consistent set of values for $\{p^{(\text{suc})}_1, p^{(\text{suc})}_2, Z(c_1), ..., Z(c_N ) \}$, i.e. solving  Eq.~(\ref{eq:ps_Kobjs}) for $K=2$ (or more simply Eq. (\ref{eq:ps_1Guy}) in the case of a single type of goods).
It can be generalized straightforwardly to $K>2$, although it may become numerically extremely expensive (see also our code, \cite{CodeLandes2015}).
The results (green crosses) presented in Figure \ref{Fig:summary_section_one_object} were obtained using the method described here.

For each agent there is a constant $Z(c_i)$ to be determined self-consistently.
This presents a technical difficulty, as for a true power-law distribution, each agent gets a different wealth and thus the number of constants to compute is $N$.

\paragraph{Staircase-like distribution of wealth (with exponent $\beta$)}

A way to tackle this difficulty is to consider a staircase-like distribution of wealth, where agents are distributed in groups with homogeneous wealth $c_g$ and where the number of agents per group is $N_g \sim \int_{c_g}^{c_{g+1}}\rho(c) \d c$, so that individual agents approximately follow a power law with exponent $\beta$. 
See Figure \ref{Fig:Various_Capital_Distro} (green crosses) to have an idea of what this modified distribution means concretely.
This kind of staircase distribution is not a true power-law, in particular because its maximum is always deterministic and finite. 
However, as we now have $1< \mathcal{N} \ll N$, we can numerically solve the $\mathcal{N}+1$ equations and thus find the exact value of $p^{(\text{suc})}$.
Of course, the value of $p^{(\text{suc})}$ found in this way perfectly matches with Monte Carlo results if and only if we use the exact same distribution of wealth and goods in the simulation. This is not surprising at all, and merely validates our iterative scheme.

\begin{figure}
\includegraphics[width=\textwidth]{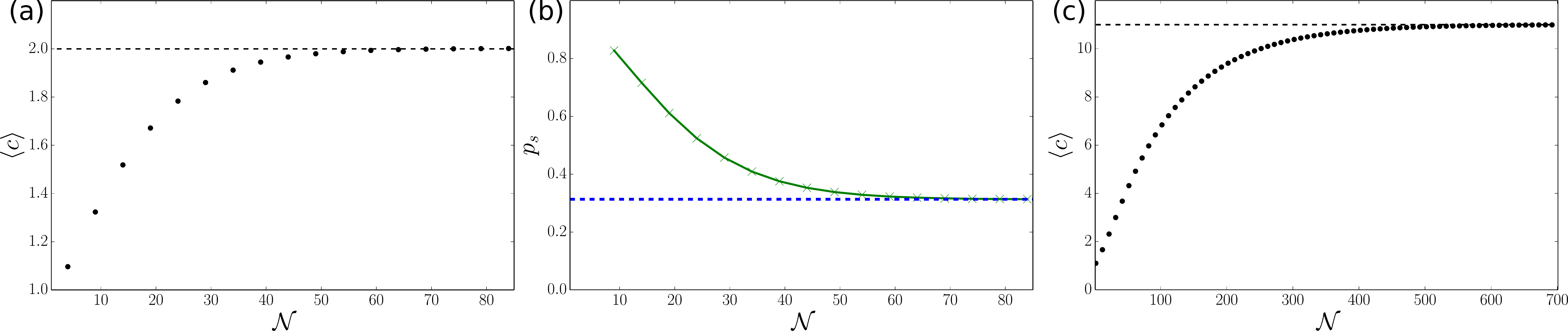}
\caption{
(a): Average wealth $ \empiri{c}$ dependence on $\mathcal{N}$ for a staircase distribution, using $\beta=2$ and $b=1.1$. Black dots: average computed (exactly) for the staircase distribution. Dashed black: expectation value for the corresponding true power law.
Convergence is reached as soon as $\mathcal{N}\approx 50$.
\label{Fig:staircase_convergence_ok}
\\
(b): Dependence on $\mathcal{N}$ of the $p^{(\text{suc})}_1$ computed from the iterative method, using a staircase-like distribution of wealth (green crosses). 
As soon as $\mathcal{N}>50$, it approaches its ``true'' value, i.e. the value obtained for a true power-law with exponent $\beta=2$ (dashed blue line).
We used $b=1.1$.
\label{Fig:staircase_convergence_ps}
\\
(c):
Average wealth $ \empiri{c}$ dependence on $\mathcal{N}$ for a staircase distribution, using $\beta=1.1$ and $b=1.1$. Black dots: average computed (exactly) for the staircase distribution. Dashed black: expectation value for the corresponding true power law.
 It takes very large $\mathcal{N}$ to converge.
\label{Fig:staircase_convergence_NOT_WORKING_ok}
}
\end{figure}

However, we note that staircase-like wealth distributions turn out to be very good approximations of true power laws, when the wealth levels $c_g$ are sufficiently refined and the number of classes $\mathcal{N}$ sufficiently large. 
In particular, using $c_g=b^g$ with a base $b \approx 1^+$, it can be seen that for large enough $\mathcal{N}$, the average wealth $\empiri{c}$ converges to a value very close to the expected one $\expect{c}$ (and no longer depends on $\mathcal{N}$).
For large $\beta$, typically $\beta \geq 1.5$, convergence is reached rather fast ($\mathcal{N}\geq 50$ is enough), and the iterative method can be used (see for instance Figure \ref{Fig:staircase_convergence_ok}a).
Under these conditions, the observables (e.g. $p^{(\text{suc})}$) have the same values for a true power law and the corresponding staircase-distribution (see Figure \ref{Fig:staircase_convergence_ps}b).
However for smaller values of $\beta$, convergence is very slow and one needs at least $\mathcal{N} > 200 $ to converge (see Figure \ref{Fig:staircase_convergence_NOT_WORKING_ok}c).  
The maximum wealth is then very large, which makes the iterative method useless for practical purposes (overflow errors arise, and the number of terms in the sums to be computed explodes exponentially, along with the computational cost).

\paragraph{The algorithm}

See also our Mathematica code, \cite{CodeLandes2015}.
The idea is the following.
We define ``old'' and ``new'' values for each of the variables $Z(c_1), ..., Z(c_\mathcal{N})$ and for $Z=\sum_{i} Z(c_i)$.
For $p^{(\text{suc})}_1, p^{(\text{suc})}_2$, we define only the ``current'' values, and have ``target'' values.
We start with an initial guess, e.g. that all the $p^{(\text{suc})}=0.5$ and $Z^{old}=1$ (which is not consistent, of course).
Then we use the new values of the $p^{(\text{suc})}$ to compute the new normalization factors. These are used to compute the ``target'' $p^{(\text{suc})}$, i.e. to know if the $p^{(\text{suc})}$ should be increased or decreased. New values of $p^{(\text{suc})}$ are again used to recompute the normalizations, and so forth.

To be more precise, here is the pseudo code we used. We start with some guess values for the $p^{(\text{suc})}$, e.g. $0.5$. Let the variables $\Delta p^{(\text{suc})}_1, \Delta p^{(\text{suc})}_2$ be set to $0.1$.
The function $P^{old}(z_1,z_2|c_i)$ uses $Z^{old}$ as normalization factor, and the current values of the $p^{(\text{suc})}$ (since the term $\lambda_k=M_k/(N p^{(\text{suc})}_k)$ is ubiquitous in the expression of $P$).
\begin{itemize}
\item Compute the new $Z(c_1), ..., Z(c_\mathcal{N})$ using the $p^{(\text{suc})}_k$: 
$Z^{new}(c_i) = \sum_{z_1} \sum_{z_2} P(z_1,z_2|c_i)$
\item Compute $Z^{new} =\sum_{i} Z^{new}(c_i)$.
\item Set $Z^{old} = Z^{old}  * Z^{new}$.
\item Compute the target values of the $p^{(\text{suc})}$: 
\begin{align}
(p^{(\text{suc})}_1)^{goal}= \sum_i^{\mathcal{N}} \left[ \frac{N_i}{N } - \frac{N_i}{N }  \lp \frac{Z^{new}}{Z^{new}(c_i)} \sum_{z_2=0}^{\floor{c_i/\pi_2}} P(z_1=\floor{(c_i-z_2 \pi_2)/\pi_1}, z_2|c_i)  \rp  \right]
\end{align}
and symmetrically for $(p^{(\text{suc})}_2)^{goal}$.
\item For each $k$, compute $\sigma_k = sign[((p^{(\text{suc})}_k)^{goal} - p^{(\text{suc})}_k) \Delta p^{(\text{suc})}_k]$.
\item For each $k$, If $\sigma_k <0 $, set $\Delta p^{(\text{suc})}_k$ to $-\Delta p^{(\text{suc})}_k/2$.
\item For each $k$, set $p^{(\text{suc})}_k$ to $p^{(\text{suc})}_k +\Delta p^{(\text{suc})}_k$. 
\item If the $p^{(\text{suc})}$ obtained is smaller than $0$, set it to $|\Delta p^{(\text{suc})}_k|$ and divide $\Delta p^{(\text{suc})}_k$ by $2$. 
\item If the $p^{(\text{suc})}$ obtained is larger than $1$, set it to $1$.
\item Loop until all the $\Delta p^{(\text{suc})}_k$ are smaller than the predefined allowed error and/or the $Z^{new}$ is close enough to $1$. 
\end{itemize}	
This algorithm converges to the true value of the $p^{(\text{suc})}$ for large enough $\mathcal{N}$. 

Typically, the $\mathcal{N}$ that is sufficient to achieve a reasonable approximated convergence can be estimated by taking a quick look at how much the $\empiri{c}(\mathcal{N})$ is close to the $\expect{c}$.
Running this algorithm at different values of $\mathcal{N}$, one can directly probe the convergence: when increasing $\mathcal{N}$ does not change the values of the $p^{(\text{suc})}$ more than the errorbar allowed, then one considers the method to have converged.
In practice, it is fairly fast to converge for large $\beta$'s, and the number of operations exponentially explodes as $\beta$ is decreased towards $1$.

\end{document}